\documentclass[preprintnumbers,amsmath,amssymb,floatfix,10pt,prd,onecolumn,
superscriptaddress,nofootinbib]{revtex4-2}
\usepackage{latexsym}
\usepackage{epsfig}
\usepackage{epstopdf}
\usepackage{graphicx}
\usepackage{amssymb}
\usepackage{amsmath}
\usepackage{amsfonts}
\usepackage{subfigure}
\usepackage{dcolumn}
\usepackage{bm}
\usepackage[normalem]{ulem}
\usepackage[dvipsnames]{xcolor}
\usepackage{hyperref}
\usepackage{color}
\usepackage{comment}
\usepackage{float}
\usepackage[utf8]{inputenc}
\hypersetup{colorlinks=true,linkcolor=blue,citecolor=red,urlcolor=magenta}
\usepackage{mathrsfs}
\usepackage{mathtools}
\usepackage{orcidlink}

\begin{document}

\title{Quantum Black Holes: Perihelion Advance, Quasi Normal Modes and Classical/ Topological Thermodynamics}

\author{Grigoris Panotopoulos}
\email{grigorios.panotopoulos@ufrontera.cl}
\affiliation{Departamento de Ciencias Físicas, Universidad de La Frontera, Casilla 54-D, 4811186 Temuco, Chile.}

\author{Francisco Tello-Ortiz \orcidlink{0000-0002-7104-5746}}
\email{francisco.tello@ufrontera.cl}
\affiliation{Departamento de Ciencias Físicas, Universidad de La Frontera, Casilla 54-D, 4811186 Temuco, Chile.}

\begin{abstract}
We report on some properties of a quantum black hole obtained recently. The correction to the Newtonian gravitational potential is proportional to a coupling $\alpha$, which is the only free parameter of the theory. We constrain the coupling using the perihelion advance, we compute the quasi-normal modes for scalar (both massless and massive) and electromagnetic perturbations. We find that all modes computed here are complex numbers characterized by a positive real part and a negative imaginary part, while both parts increase with the mass of the test scalar field. Also thermodynamics properties are investigated from the classical and topological point of view. In this regard, the quantum black hole exhibits the same behavior as the classical Reissner-Nordstr\"om space-time, that is, it presents stable/unstable branches in the Gibbs potential, one generating point and a topological charge $W=0$. 
\end{abstract}

\maketitle

\section{Introduction}

Ideal black holes (BHs) are supposed to be isolated objects. Realistic BHs of Nature, however, are in constant interaction with their environment. We may think, for instance, matter accretion onto a BH from its donor in binaries. When a BH is perturbed due to a certain interaction, the geometry of space-time undergoes damped oscillations. How a system responds to small perturbations as well as normal modes of oscillating systems have always been important topics in physics. Regarding BH physics in particular, the work of \cite{regge} long time ago marked the birth of BH perturbation theory, and later on it was extended by other people \cite{zerilli1,zerilli2,zerilli3,moncrief,teukolsky}. Nowadays the state-of-the art in BH physics and perturbations is nicely summarized in the comprehensive review of Chandrasekhar's monograph \cite{monograph}. The information on how a given BH relaxes after the perturbation has been applied is encoded into the quasi-normal (QN) frequencies. The latter are complex numbers, with a non-vanishing imaginary part, that depend on the details of the background geometry as well as the spin of the propagating field at hand (scalar, Dirac, vector (electromagnetic), tensor (gravitational)), but they do not depend on the initial conditions. Therefore, QN modes (QNMs) carry unique information about black hole physics. Black hole perturbation theory and QNMs of black holes are relevant during the ringdown phase of binaries, in which after the merging of two black holes a new, distorted object is formed, while at the same time the geometry of space-time undergoes damped oscillations due to the emission of gravitational waves. For a review on QNMs of BHs see e.g. \cite{Berti:2009kk, Konoplya:2011qq}.

\smallskip

Another way to test the stability of BHs solutions, besides the QNMs approach, is the thermodynamics framework. The thermodynamics setting of gravity and vice versa, was put forward by seminal articles by Bardeen, Carter and Hawking \cite{Bardeen:1973gs}, Bekenstein \cite{Bekenstein:1973ur}, Hawking \cite{Hawking:1975vcx}, Gibbons and Hawking \cite{Gibbons:1976ue} and York \cite{York:1986it}. In these articles, the BH thermodynamics laws were developed, yielding  a very important interpretation, such as the BH temperature as surface gravity \cite{Hawking:1975vcx, Gibbons:1976ue} and the BH entropy as surface area \cite{Bekenstein:1973ur}. These results were obtained within the framework of General Relativity (GR). Later, Jacobson was able to obtain Einstein field equations from a pure thermodynamic description \cite{Jacobson:1995ab}. Although this methodology was derived in the 3+1 dimensions, Myers and Simon in \cite{Myers:1988ze} extended its usage to Lovelock gravity \cite{Lovelock:1971yv}. Furthermore, the idea that the mass of an AdS BH should be understood as the enthalpy of the spacetime marks a significant advancement in the field. This concept arose from geometric derivations of the Smarr formula for AdS BHs, which indicated that the cosmological constant could be treated as a thermodynamic variable, analogous to pressure in the first law of thermodynamics \cite{Kastor:2009wy,Altamirano:2014tva}. A recently introduced concept in BH thermodynamics is the notion of thermodynamic topology. Since its inception \cite{Wei:2021vdx,Wei:2022dzw}, this idea has been further developed in several notable works \cite{Zhu:2024zcl,Liu:2024lbi,Liu:2024oas,Liu:2024soc,EslamPanah:2024fls,Bakopoulos:2024ogt,Bhattacharya:2024bjp} (and references therein). Within this framework, BH solutions are interpreted as topological defects in their thermodynamic space. Both local and global topology can be studied by calculating the winding numbers at these defects. BHs are then classified according to their total topological charge. Moreover, BH thermal stability is linked to the sign of its winding number. A central feature of thermodynamic topology is the association of topological defects with their corresponding topological charges.

\smallskip

Based on the ground of these schemes to theoretically test BH stability, the main aim of this article is to analyze whether the seminal loop quantum cosmology (LQC) static BH provided in \cite{Lewandowski:2022zce}, is stable under mechanical perturbations and the thermodynamics setting. In this regard, interesting articles have been devoted to elucidate the main aspect of this corrected quantum Schwarzschild BH under different approaches. For example, \cite{Yang:2022btw} was the first to investigate the quasinormal modes of quantum-corrected black holes, \cite{Shao:2023qlt} initially proposed the concept of a quantum black hole with a cosmological constant, and \cite{Lin:2024flv} was the first to derive such a quantum black hole solving the quantum-corrected dynamical equations.
What is more, in \cite{Skvortsova:2024atk,Gong:2023ghh} a QNMs study has been carried out, in \cite{Zhao:2024elr} and \cite{Ye:2023qks} some optical features have been considered, such as gravitational lensing and the shadow, in \cite{Wang:2024jtp} and extension of this model including the cosmological and thermodynamics related properties were studied. In addition, novel wormhole solutions have been found within this framework \cite{Mazharimousavi:2024nih}. The main a major difference between the present study and these antecedents is that in all these cases, the value of the quantum parameter $\gamma$ (or $\alpha$) has been fixed to its theoretical value reported in \cite{Meissner:2004ju,Domagala:2004jt}. This represents a limitation, in studies, because we force the results to satisfy only theoretic restrictions\footnote{In \cite{Zhao:2024elr}, the authors constrained the quantum parameter using the EHT observational data. This represents a more realistic scenario on the possibility of having a quantum BH solution.}. So, in this work, we first bound the quantum parameter using the perihelion advance phenomenology, and from this analysis we chose the value for this important parameter. On the other hand, for the QNMs analysis, only one value of overtone number $n$ has been used. As far as we know, the overtone number $n$ measures how the different QNMs contribute to the decay and oscillation of the BH perturbed field, reflecting the complex dynamics of space-time around these extreme objects. Furthermore, we consider massless and massive scalar field and the electromagnetic one. From the point of view of thermodynamics analysis, we perform the topological approach. However, to properly applied this off shell protocol, one needs to ensure that the usual thermodynamics first law is valid, that is, the BH horizon mass $M$ coincides with the BH energy, and Hawking temperature coincides with the temperature obtained from the first law. Clearly, this is not the case because the entropy of this LQC static BH does not meet the holographic principle (the Bekenstein entropy \cite{Bekenstein:1973ur}). This is problematic since the topological approach must coincide with the classical thermodynamics description once the parameter $\tau$ is recognized as the inverse of the BH temperature. In this regard, we have ``regularized'' the first law to get a temperature matching Hawking expression. 

\smallskip

The article is organized as follows: Sect. \ref{sec2} presents a short review about the quantum BH. In Sect. \ref{sec3}, the quantum parameter $\alpha$ is bounded using the perihelion advance phenomenology. Also, it is discussed how this quantum correction is affecting the Newtonian potential. Sects. \ref{sec4} and \ref{sec5} present the stability analysis via QNMs using the WKB scheme and the thermodynamics study using the classical and topological approaches, respectively. Finally, Sect. \ref{sec6} concludes the work. Throughout this article the mostly positive signature has been used $\{-;+;+;+\}$, and geometrical units where $c=G=1$, then $\kappa=8\pi$, have been adopted.

\section{Schwarzschild-like black hole in LQG}\label{sec2}

In this section, we briefly review the main aspect of quantum BH provided in \cite{Lewandowski:2022zce}. This space-time is being described by the following line element 
\begin{equation}
\mathrm{d} s^2=-f(r) \mathrm{d} t^2+\frac{1}{f(r)} \mathrm{d} r^2+r^2 \mathrm{~d} \theta^2+r^2 \sin ^2 \theta \mathrm{~d} \phi^2,
\end{equation}
with 
\begin{equation}\label{blacken}
f(r)=1-\frac{2 M}{r}+\frac{\alpha M^2}{r^4},
\end{equation}
where $\alpha=16\sqrt{3}\pi \gamma^{3}$, with $\gamma$ being the Barbero-Immirzi parameter in LQG \cite{BarberoG:1994eia,Meissner:2004ju,Domagala:2004jt}. The coupling $\alpha$ has units of square length and $\gamma$ is a dimensionless parameter. Some studies such as \cite{Meissner:2004ju,Domagala:2004jt} suggest that $\gamma\approx 0.2375$, therefore $\alpha \approx 1.1663$. 
The causal structure of this space-time, is quite similar to the Reissner-Nordstr\"om BH, that is, it has a Cauchy and event horizon. Despite the fact that the polynomial expression is of fourth order, given the constraints imposed on $\alpha$, the polynomial has two complex roots, while the remaining two real roots represent the horizons. From the blacken function (\ref{blacken}) the horizon mass of this BH reads
\begin{equation}
M(r_{+})=\frac{r^{2}_{+}\left(r_{+}\pm\sqrt{r^{2}_{+}-\alpha}\right)}{\alpha}.
\end{equation}
with $r_-, r_+$ being the inner and outer horizon, respectively.

It is worth mentioning that from this expression, the upper branch should be discarded because it does not reproduce the adequate Schwarzschild limit when $\alpha\rightarrow 0$. Hence, the correct branch is the lower one. This branch defines a minimum values for the event horizon $r_{+}$ given by \cite{Lewandowski:2022zce}
\begin{equation}
    r_{+}\geq \sqrt{\alpha}=r^{\text{min}}_{+},
\end{equation}
then the minimum horizon mass is 
\begin{equation}
    M_{min}(r_{+})=\frac{4\sqrt{\alpha}}{3\sqrt{3}}.
\end{equation}
So, it is clear that the minimum horizon radius and horizon mass, strongly depend on the values of the parameter $\alpha$. In this study, we are going to bound $\alpha$ (consequently $\gamma$) using the perihelion advance phenomenology (see next section \ref{sec3}), and in view of this result, we are going to study the stability of this model using the QNMs and thermodynamic approaches.

\section{Perihelion advance for Quantum corrected BH}\label{sec3}

In this section, we present the first part of the analysis performed in this work, namely how to constrain the parameter $\alpha$ using observational data coming from the perihelion advance of the planet Mercury around the Sun.

\smallskip

In the non-relativistic limit, the following relation holds \cite{Wald, Landau}:
\begin{equation}\label{newton}
2 \Phi(r) + 1 = f(r)=g_{00}(r) = 1-\frac{2 M}{r} + \frac{\alpha M^2}{r^4}
\end{equation}
with $\Phi(r)$ being the gravitational potential. Thus, in the theory of gravity with quantum corrections, the total gravitational potential consists of two terms
\begin{equation}
\Phi(r)=-\frac{M}{r} + \frac{\alpha M^2}{2 r^4}
\end{equation}
while the gravitational potential energy, $V(r)$, is simply given by $V(r) = m \Phi(r)$, with $m$ being the mass of a test particle in the fixed gravitational background. Therefore, there is a single perturbing potential beyond the Newtonian one, namely the one due to quantum corrections. 

\smallskip

A generic and useful expression for the perihelion advance, $\Delta \theta_p$, due to any perturbative potential energy, $V(r)$, beyond the Newtonian one, is found to be (setting $G=1=c$) \cite{Adkins:2007et}
\begin{equation}
\Delta \theta_p = \frac{-2 L}{m M e^2} \int_{-1}^{+1} \frac{dz z}{\sqrt{1-z^2}} \frac{dV}{dz}
\end{equation}
where $L = a (1-e^2)$, the perturbing potential energy is evaluated at $r = L/(1 + e z)$, and $e$ and $a$ are the eccentricity and the semi-major axis of the orbit, respectively. Let us mention that, as was indicated in \cite{Zakharov:2018omt}, the above expression is still valid in modified theories of gravity. The study of the motion of test particles in a given gravitational background (geodesic equations via the Christoffel symbols) remains the same in all metric theories of gravity, irrespectively of the underlying theory. The general expression for the precession angle in terms of the perturbing potential has been derived considering the orbit $u(\theta)$, where $u = 1/r$, and this expression does not depend on the underlying theory of gravity.

\smallskip

In the present work, clearly there are in total two contributions, namely (i) the Newtonian potential ($\Delta \theta_p(G R)$) and (ii) the new term coming from quantum corrections ($\Delta \theta_p(Q G)$). The GR contribution is computed to be (setting $G = 1 = c$) \cite{Adkins:2007et}
\begin{equation}
\Delta \theta_p(G R) = \frac{6 M \pi}{L} 
\end{equation}
while the other contribution is found to be
\begin{equation}
\Delta \theta_p (Q G)= \frac{-3 \alpha \pi M (4+e^2)}{2 L^3} 
\end{equation}
The total contribution is given by
\begin{equation}
\Delta \theta_p = \Delta \theta_p(G R) + \Delta \theta_p (Q G).
\end{equation}
From the observational point of view, here, we shall use the precession angle of the planet Mercury \cite{Clifton:2020xhc, Pitjeva:2013xxa}
\begin{equation}
\Delta \theta_p - \Delta \theta_p (G R) = (-0.002 \pm 0.003)" \textrm{per century}.
\end{equation}

\smallskip

Finally, regarding the details of the orbit, in the case of Mercury we use the following numerical values \cite{Clifton:2020xhc}
\begin{eqnarray}
M & = & 1.99 \times 10^{30} kg \\
a & = & 5.79 \times 10^7 km  \\
e & = & 0.20563.
\end{eqnarray}


\begin{figure*}[ht!]
\centering
\includegraphics[scale=0.9]{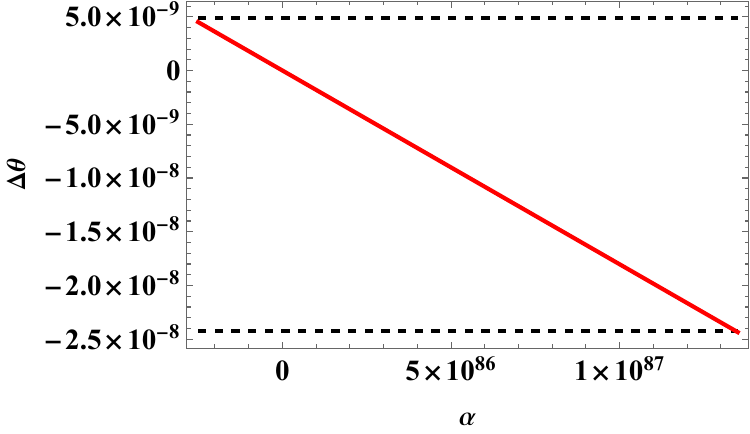} 
\caption{
Observational bound on the parameter $\alpha$ coming from the perihelion advance of Mercury. The solid line corresponds to the prediction of the model, while the strip corresponds to the observationally allowed range.
}
\label{fig:1} 	
\end{figure*}


In Fig. \ref{fig:1} we show the perihelion advance versus $\alpha$, both the theoretical prediction (solid red curve) and the allowed region (horizontal strip) according to observations. It should be mentioned that the parameter $\alpha$ has dimensions $(length)^2$, although in the plot it is dimensionless, and it is to be understood that it is expressed in units of $l_{pl} \sim 10^{-33} \: cm$, with $l_{pl}$ being the Planck length. In order to demonstrate that the correction is indeed small, we shall compare it to the usual Newtonian term of the well-known Schwarzschild geometry, i.e.
\begin{equation}\label{correct}
\frac{\alpha \: (GM)^2/r^4}{GM/r} \sim 10^{-10} \ll 1
\end{equation}
for $\alpha \sim 10^{87} \: l_{pl}^2$, for distances up to the size of the orbit of Mercury, $r \sim a \sim 10^{45} l_{pl}$, and for a solar mass $M \sim 10^{38} \: m_{pl}$, with $m_{pl}$ being the Planck mass.

Notice that the value of the coupling $\alpha$ (in natural units $c=G=\hbar=1$) proposed in \cite{Ye:2023qks,Wang:2024jtp}
\begin{equation}
\alpha = 16 \sqrt{3} \pi \gamma^3, \, \, \, \, \, \gamma \approx 0.2375
\end{equation}
falls within the bound obtained here. As it is of order one, and therefore much lower than $10^{87}$, it would be practically zero in the plot shown here.

\section{Stability via quasi normal modes}\label{sec4}

We study the propagation of test particles in a fixed gravitational background. First, regarding scalar perturbations, we consider a minimally coupled scalar field, $\Psi$, in the above background with lapse function $f(r)$. The equation of motion is the standard Klein-Gordon equation
\begin{equation}
\frac{1}{\sqrt{-g}} \partial_\mu (\sqrt{-g} g^{\mu \nu} \partial_\nu \Psi) = m^2 \Psi,
\end{equation}
with $m$ being the mass of the scalar field. Using as usual the ansatz $\Psi(t,r,\theta, \phi)=e^{-i \omega t} R(r) Y_l^m(\theta, \phi)$, where $Y_l^m$ are the standard spherical harmonics, we obtain the radial equation
\begin{equation} \label{radial}
R'' + \left(\frac{f'}{f}+\frac{2}{r}\right) R' + \left(\frac{\omega^2}{f^2}-\frac{l (l+1)}{r^2 f} - \frac{m^2}{f}\right) R = 0.
\end{equation}
To see the potential that the scalar field feels we define new variables as follows
\begin{eqnarray}
R & = & \frac{\psi}{\sqrt{r}}, \\
x & = & \int \frac{dr}{f(r)},
\end{eqnarray}
where $x$ is the so called tortoise coordinate. We recast the equation for the radial part into a Schr{\"o}dinger-like equation of the form \cite{Ferrari:2007dd, Berti:2009kk, Konoplya:2011qq}
\begin{equation}
\frac{d^2 \psi}{dx^2} + [ \omega^2 - V(x) ] \psi = 0,
\end{equation}
with $\omega$ being the QN frequencies, and we obtain for the effective potential barrier the expression
\begin{equation}
V_{sc}(r) = f(r) \: \left( m^2 + \frac{l (l+1)}{r}+\frac{f'(r)}{r} \right),
\end{equation}
with $l=0,1,2,...$ being the angular degree, and the prime denotes differentiation with respect to $r$.

\smallskip

Regarding electromagnetic perturbations, the starting point is Maxwell equations. Following similar steps as before for the case of the scalar field, one finally obtains a a Schr{\"o}dinger-like equation, where this time the effective potential barrier is given by
\begin{equation}
V_{EM}(r) = f(r) \: \frac{l (l+1)}{r^2}.
\end{equation}

Therefore, the potential barrier for massless perturbations of spin $s=0,1$ is found to be \cite{Berti:2009kk}
\begin{equation}
V_{m=0}(r) = f(r)
\Bigg[ 
\frac{\ell(\ell + 1)}{r^{2}} + \frac{f'(r)}{r} (1-s^2),
\Bigg],
\label{poten}
\end{equation}
while for massive scalar perturbations the potential barrier is given by
\begin{equation}
V_m(r) = f(r)
\Bigg[ m^2 +
\frac{ \ell(\ell + 1)}{r^{2}} + \frac{f'(r)}{r}
\Bigg].
\label{poten}
\end{equation}

Finally, the wave equation must be supplemented by the following boundary conditions \cite{Ferrari:2007dd, Konoplya:2011qq}
\begin{equation}
\psi \rightarrow \: \exp( i \omega x), \; \; \; \; \; \; x \rightarrow - \infty \ ,
\end{equation}
\begin{equation}
\psi \rightarrow \: \exp(-i \omega x), \; \; \; \; \; \; x \rightarrow  \infty \ .
\end{equation}


\begin{table}[H]
\centering
\caption{Quasinormal frequencies for massless scalar perturbations (varying $\ell$ and $n$) fixing $M=1,\alpha=0.02$ for the model considered in this work. 
}
{
\begin{tabular}{c|ccccc} 
\toprule
$n$  &  $\ell=1$ & $\ell=2$ & $\ell=3$ & $\ell=4$ & $\ell=5$
\\ \colrule
\hline
    0 &  0.293028-0.097691 I  & 0.483827-0.096691 I & 0.675621-0.096427 I & 0.867740-0.096319 I & 1.060010-0.096264 I  \\
    1 &      			  	  & 0.464086-0.295382 I & 0.660964-0.292055 I & 0.856163-0.290651 I & 1.050460-0.289932 I  \\
    2 &  				  	  &  			        & 0.633954-0.495588 I & 0.834096-0.489924 I & 1.031960-0.486958 I \\
    3 &                       &                     &                     & 0.803672-0.696880 I & 1.005700-0.689348 I \\
    4 &                       &                     &                     &                     & 0.973387-0.898706 I \\
\botrule
\hline
\end{tabular} 
\label{table:First set}
}
\end{table}


\begin{table}[H]
\centering
\caption{Quasinormal frequencies for massive scalar perturbations (varying $\ell$ and $n$) fixing $M=1, \alpha=0.02, m=0.1$ for the model considered in this work. 
}
{
\begin{tabular}{c|ccccc} 
\toprule
$n$  &  $\ell=1$ & $\ell=2$ & $\ell=3$ & $\ell=4$ & $\ell=5$
\\ \colrule
\hline
    0 & 0.297503 - 0.0949935 I & 0.486984 - 0.095609 I & 0.677979 - 0.095858 I & 0.869609 - 0.095970 I & 1.061550 - 0.096029 I \\
    1 &                        & 0.465707 - 0.293046 I & 0.662672 - 0.290615 I & 0.857706 - 0.289708 I & 1.051820 - 0.289274 I  \\
    2 &                        &                       & 0.634759 - 0.493869 I & 0.835114 - 0.488643 I & 1.033000 - 0.486003 I \\
    3 &                        &                       &                       & 0.804156 - 0.695551 I & 1.006370 - 0.688258 I  \\
    4 &                        &                       &                       &                       & 0.973713 - 0.897632 I \\
\botrule
\hline
\end{tabular} 
\label{table:Second set}
}
\end{table}


\begin{table}[H]
\centering
\caption{Quasinormal frequencies for massive scalar perturbations (varying $\ell$ and $n$) fixing $M=1,\alpha=0.02, m=0.2$ for the model considered in this work. 
}
{
\begin{tabular}{c|ccccc} 
\toprule
$n$  &  $\ell=1$ & $\ell=2$ & $\ell=3$ & $\ell=4$ & $\ell=5$
\\ \colrule
\hline
    0 & 0.311031 - 0.086646 I & 0.496498 - 0.092332 I & 0.685071 - 0.094143 I & 0.875223 - 0.094922 I & 1.066180 - 0.095324 I \\
    1 &      			  	  & 0.470532 - 0.285974 I & 0.667797 - 0.286268 I & 0.862338 - 0.286868 I & 1.055900 - 0.287298 I \\
    2 &  				  	  &  			          & 0.637150 - 0.488690 I & 0.838165 - 0.484785 I & 1.036120 - 0.483131 I \\
    3 &                       &                       &                       & 0.805592 - 0.691557 I & 1.008370 - 0.684981 I \\
    4 &                       &                       &                       &                       & 0.974682 - 0.894404 I \\
\botrule
\hline
\end{tabular} 
\label{table:Third set}
}
\end{table}


\begin{table}[H]
\centering
\caption{Quasinormal frequencies for electromagnetic perturbations (varying $\ell$ and $n$) fixing $M=1,\alpha=0.02$ for the model considered in this work. 
}
{
\begin{tabular}{c|ccccc} 
\toprule
$n$  &  $\ell=1$ & $\ell=2$ & $\ell=3$ & $\ell=4$ & $\ell=5$
\\ \colrule
\hline
    0 & 0.248343-0.092579 I  & 0.457792-0.094937 I & 0.657161-0.095544 I & 0.853426-0.095787 I  & 1.048310-0.095909 I \\
    1 &      			  	 & 0.436796-0.290485 I & 0.642042-0.289499 I & 0.841630-0.289090 I  & 1.038650-0.288882 I \\
    2 &  				  	 &  			       & 0.614169-0.491642 I & 0.819142-0.487437 I  & 1.019920-0.485259 I \\
    3 &                      &                     &                     & 0.788134-0.693636 I  & 0.993321-0.687076 I \\
    4 &                      &                     &                     &                      & 0.960588-0.895967 I \\
\botrule
\hline
\end{tabular} 
\label{table:Fourth set}
}
\end{table}


Before we proceed with our analysis a comment is in order here. To study the propagation of test particles we are using the usual Klein-Gordon and Maxwell equations without any modifications, despite the fact the BH solution discussed in the present work is a quantum corrected one. This may be justified as follows: Both the Klein-Gordon and the Maxwell equations are wave equations compatible with relativity and quantum mechanics. In any metric theory of gravity, to take into account gravitational effects, as is well known from standard textbooks, one needs to make the following two replacements
\begin{equation}
\eta_{\mu \nu} \rightarrow g_{\mu \nu}, \; \; \; \; \; \partial_\mu \rightarrow D_{\mu},
\end{equation}
where $\eta_{\mu \nu}$ is the Minkowski metric tensor of flat space-time, while $D_{\mu}$ is the covariant derivative. The details of the gravitational theory and of how the given gravitational background is obtained are irrelevant. Therefore, within Einstein's classical theory of gravitation the fixed gravitational background is the usual Schwarzschild geometry, whereas when quantum corrections are included the new gravitational background is the one discussed here.

In this study, we have computed the QNMs numerically adopting the extensively used semi-analytical WKB method \cite{Schutz:1985km,iyer:1986np}. The numerical values of the frequencies are shown in the Tables \ref{table:First set}, \ref{table:Second set}, \ref{table:Third set} (scalar perturbations) and \ref{table:Fourth set} (electromagnetic perturbations). We observe that all frequencies computed here are characterized by a positive real part and a negative imaginary part. For all modes, for a given angular degree, both the real part and the imaginary part decrease with the overtone number. What is more, for a given $n$, the real part and the absolute value of the imaginary part increases with $l$, whereas the behavior of the imaginary part depends on the case. In particular, in the case of electromagnetic perturbations as well as in the case of massive scalar perturbations $m=0.1$, the absolute value of the imaginary part of the excited modes ($n=1,2,...$) decreases with $l$, whereas that of the fundamental mode ($n=0$) increases with $l$. In the case of massless scalar perturbations the imaginary part of all modes decrease with the angular degree, while for the massive scalar perturbations $m=0.2$ it is observed the following pattern: For the fundamental and first excited modes the imaginary part increases with $l$, whereas for the higher excited modes the imaginary part decreases with $l$.


\section{Thermodynamics analysis}\label{sec5}

In this section, we perform a thermodynamical analysis using the classical approach, that is, by computing the usual thermodynamic properties such as the Hawking temperature, Gibbs free energy, and using a novel technique based on topological charges. In the following subsection, we tackle the classical scheme, by correcting the first thermodynamic law in order to obtain a well-posed description. This step is necessary because the horizon mass of the quantum BH percolates into the energy-momentum tensor component, leading to an incompatibility of the BH temperature obtaining from the first thermodynamic law and Hawking approach. This is a usual problem when BH entropy is not matching Bekenstein-Hawking formula. Next, we proceed characterizing the thermodynamic description using Duan's \cite{Duan:1984ws} scheme, recently translated into the BH arena \cite{Wei:2021vdx,Wei:2022dzw}.

\subsection{Classical approach}

It is well known that when the BH mass $M$ percolates into the components of the energy-momentum tensor, the first thermodynamic law ceases to be valid \cite{Rodrigues:2022qdp,Singh:2020xju}
\begin{equation}\label{TFL}
    dM\neq TdS.
\end{equation}
Furthermore, the BH entropy does not meet the Bekenstein-Hawking form, leading it to a discrepancy between the Hawking temperature $T_{H}$ and the one obtained from the first thermodynamics law $T_{h}$
\begin{equation}
    T_{H}=\frac{f'(r_{+})}{4\pi}\neq T_{h}=\frac{\partial M}{\partial S}.
\end{equation}
One way to reconcile at least the standard definitions for the entropy $S$ and temperature $T$, is by modifying the thermodynamics first law (\ref{TFL}) introduction a correction term $\mathcal{W}(M,r_{+})$ as follows\footnote{Notice that the parameter $\alpha$ has not be considered in the first law. This is so because $\alpha$ is taken to be constant, then $d\alpha=0$.}
\begin{equation}\label{TFL1}
    \mathcal{W}dM=TdS\Rightarrow d\mathcal{M}=TdS,
\end{equation}
where the correction factor $\mathcal{W}(M,r_{+})$ is defined by
\begin{equation}
    \mathcal{W}(M,r_{+})=1+4\pi \int^{+\infty}_{r_{+}}r^{2}\frac{\partial T^{0}_{0}}{\partial M}dr,
\end{equation}
with $T^{0}_{0}$ the temporal component of the energy-momentum tensor.
In this way, $T=T_{H}$, then $T_{H}=\mathcal{W}T_{h}$, and $S=\pi r^{2}_{+}$, where all new contributions carrying out by the black hole are included in $r_{+}$. However, in doing this, the black hole mass $M$ cannot be recognized as the energy $E$ of the system, instead of it one has
\begin{equation}\label{energy}
    dE=\mathcal{W}(M,r_{+})dM.
\end{equation}
For the present case, the horizon mass $M$ and the factor $\mathcal{W}$ are given by
\begin{equation}\label{horizonmass}
    M(r_{+})=\frac{r^{2}_{+}\left(r_{+}-\sqrt{r^{2}_{+}-\alpha}\right)}{\alpha},
\end{equation}
and 
\begin{equation}\label{factor}
   \mathcal{W}= \frac{\sqrt{r^{2}_{+}-\alpha}}{r_{+}},
\end{equation}
respectively. In obtaining the factor $\mathcal{W}$ we have used $T^{0}_{0}=-3M^{2}\alpha/r^{6}$ and then replaced (\ref{horizonmass}). It should be pointed out that the horizon mass (\ref{horizonmass}) corresponds to the smaller branch, leading it a well-defined BH mass at large enough distances. Now, using (\ref{factor}) and (\ref{horizonmass}) along with (\ref{energy}), one gets the following expression for the energy $E$
\begin{equation}\label{energyhorizon}
    E(r_{+})=\frac{1}{\alpha}\left[2\alpha r_{+} -r^{3}_{+}+(r^{2}_{+}-\alpha)^{3/2}\right].
\end{equation}
It is worth mentioning that, although the space-time is asymptotically flat\footnote{In this case, the parameter $M$ coincides with the ADM mass. Of, course one needs to take into account the lower branch.}, its horizon mass cannot be considered as the energy of the system if one wants to keep a healthy relation between the Hawking temperature and the one coming from the first thermodynamic law. Moreover, it is necessary to preserve the Bekenstein-Hawking entropy formula. In this case, the quantum correction enter via the energy and temperature (see below).

\smallskip

\begin{figure*}[h!]
\centering
\includegraphics[width=0.45\textwidth]{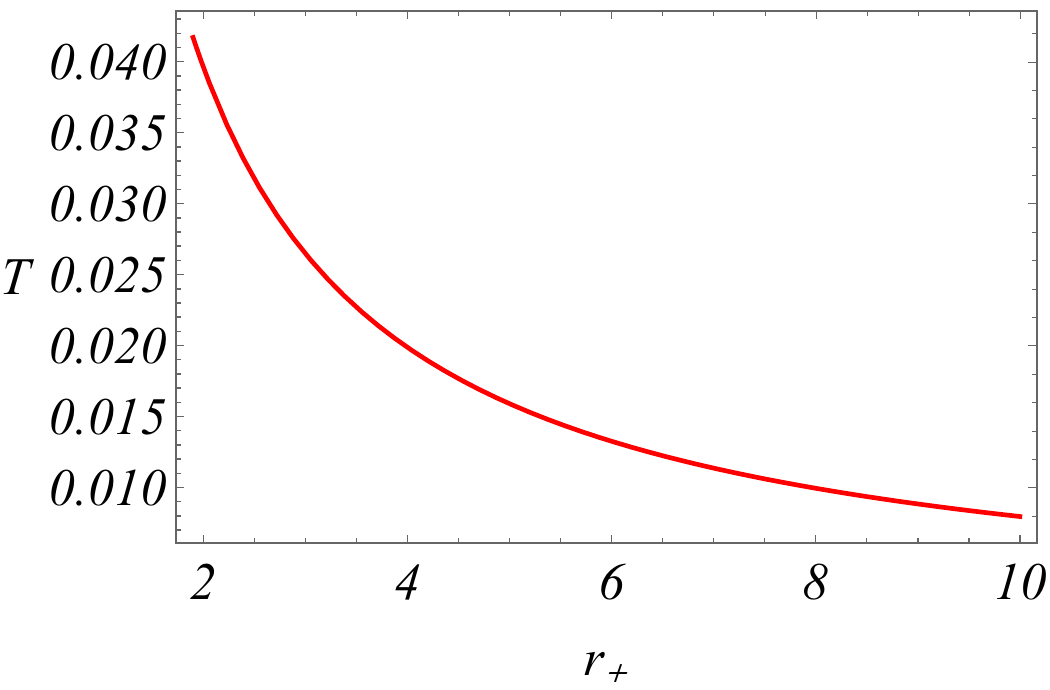}   \
\includegraphics[width=0.42\textwidth]{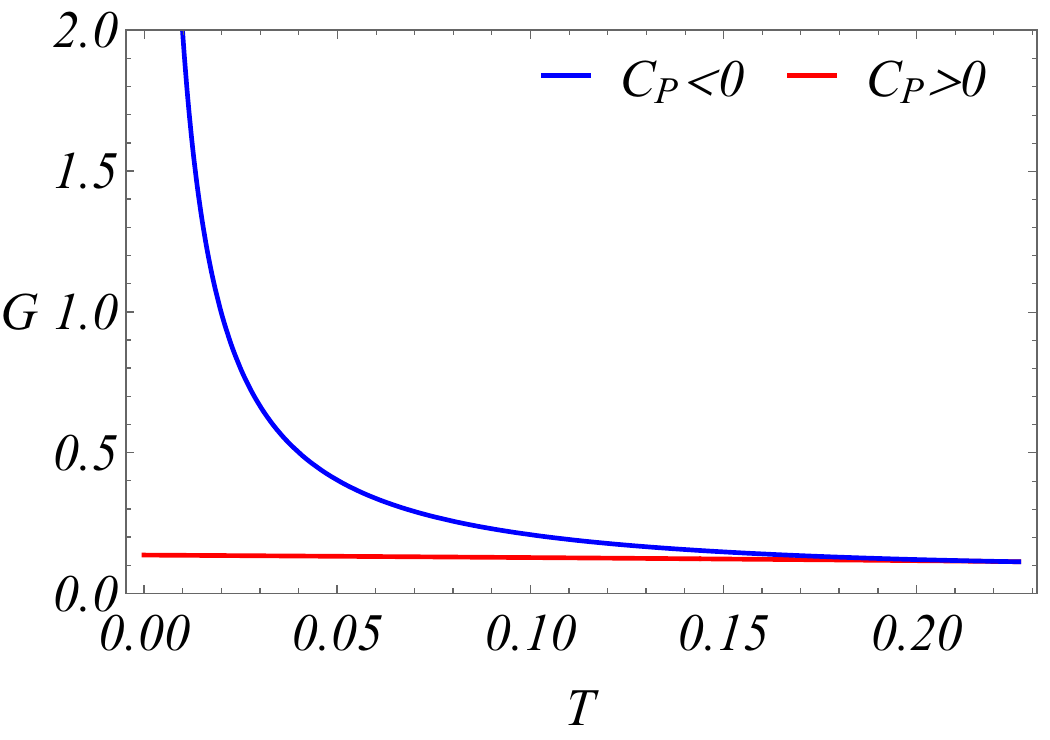}    
\caption{The trend of the BH temperature versus the BH horizon (left panel). The behavior of the Gibbs free energy against the temperature (right panel). As can be observed there a large BH (blue branch) and a small BH (red branch). The first one correspond to an unstable branch and the second one to a stable branch. This fact is coming from the signature acquire by the heat capacity at constant pressure. To build these plots, we used $M=1$ and $\alpha=0.02$. 
\label{fig}
}
\end{figure*}

Next, the temperature is given by the following expression
\begin{equation}
    T(r_{+})=\frac{f(r_{+})}{4\pi}=\frac{1}{2\pi}\left[\frac{2}{r_{+}}+\frac{3\left(\sqrt{r^{2}_{+}-\alpha}-r_{+}\right)}{\alpha}\right].
\end{equation}
As can be observed from the left panel of Fig. \ref{fig}, the temperature is positive defined as should be. Classically speaking, this BH is not presenting first order phase transitions, instead one can discriminate between an unstable BH (large BH) and stable BH (small BH). This fact is evident from the behavior of the Gibbs potential against the temperature exhibited in the right panel of Fig. \ref{fig}. Here one has a branch where the heat capacity at constant pressure is negative (blue line), this corresponds to the unstable branch and the stable branch where the heat capacity is positive (red line). Therefore, the classical thermodynamic approach reveals that this model has a large and small BH branches. It is remarkably to note that this quantum BH, behaves thermodynamically speaking in a similar way as the classical well-known Reissner-Nordstr\"om BH \cite{Altamirano:2014tva}.

\subsection{Topological approach}

To further understand the thermodynamics behavior of this quantum BH model, we are going to explore its topological interpretation following Duan's procedure \cite{Duan:1984ws}. Here, the main concept related to defects is the topological charge. In order to analyze the thermodynamic topology of
a BH, we compute the topological charge and use it to identify the topological classes. The specific method we
employ to calculate the topological charge is known as Duan’s $\phi$ mapping technique \cite{Wei:2021vdx,Wei:2022dzw}. 
To do so, we need to construct the $\phi$ mapping as follows \cite{Wei:2021vdx,Wei:2022dzw}
\begin{equation}\label{phi}
\phi=\left(\phi^r, \phi^{\theta}\right)=\left(\frac{\partial \mathcal{F}}{\partial r_{+}},-\cot \theta \csc \theta\right),
\end{equation}
where $\mathcal{F}$ is the off-shell free energy, given by
\begin{equation}
    \mathcal{F}=M-\frac{S}{\tau}.
\end{equation}
Here, the $\phi^{\theta}$  component, the trigonometric function is chosen so that one zero point of the vector field can always be
found at $\theta=\frac{\pi}{2}$. The other zero point can also be found by simply solving the equation $\phi^{r}=0$, which always results
in $\tau=\frac{1}{T}$. The basic topological property associated with the zero point or topological defect of a field is its winding
number or topological charge. In this work, we use Duan’s $\phi$ mapping technique \cite{Duan:1984ws} to calculate the winding
number. To find the topological charge we first determine the unit vector n of the field in Eq. (\ref{phi}), which are
\begin{equation}\label{unit}
\begin{aligned}
n^1 & =\frac{\phi^r}{\sqrt{\left(\phi^r\right)^2+\left(\phi^{\theta}\right)^2}}, \\
n^2 & =\frac{\phi^{\theta}}{\sqrt{\left(\phi^r\right)^2+\left(\phi^{\theta}\right)^2}}.
\end{aligned}
\end{equation}
For the vector field, a topological current can be constructed in the coordinate space $x^{\mu}=(\tau,r_{+},\theta)$ as follows \cite{Duan:1984ws}
\begin{equation}\label{current}
j^\mu=\frac{1}{2 \pi} \epsilon^{\mu \nu \rho} \epsilon_{a b} \partial_\nu n^a \partial_\rho n^b.
\end{equation}
The fundamental conditions that have to be fulfilled by the normalized vector $n^{a}$ are
\begin{equation}
n^a n_a=1 \quad \text { and } \quad n^a \partial_\nu n^a=0 .
\end{equation}
The current given in Eq. (\ref{current}), is a conserved quantity, which can be verified by applying the current conservation
law
\begin{equation}
\partial_\mu j^\mu=0 .
\end{equation}
where we use the following properties of Jacobi tensor
\begin{equation}
\epsilon^{a b} J^\mu\left(\frac{\phi}{x}\right)=\epsilon^{\mu \nu \rho} \partial_\nu \phi^a \partial_\rho \phi^b.
\end{equation}
the topological charge $W$ is related to the $0^{\text{th}}$ component of the current density of the topological current
through the following relation
\begin{equation}
W=\int_{\Sigma} j^0 d^2 x=\sum_{i=1}^N \beta_i \eta_i=\sum_{i=1}^N w_i,
\end{equation}
where $w_i$ is the winding number around the zero point. Also, $\beta_{i}$ y $\eta_{i}$ are the Hopf index and the Brouwer degree,
respectively. The detailed derivation of the above formula can be referred to \cite{Wei:2021vdx,Wei:2022dzw}.

\smallskip

Hence, the topological charge is also nonzero only at the zero points of the vector field. To find the exact zero point
where the topological charge is to be calculated, we plot the unit vector field n and find out the zero point at which
it diverges. The zero point always turns out to be $(\frac{1}{T},\frac{\pi}{2})$. Next, a contour is chosen around each zero point and is
parametrized as
\begin{equation}
\left\{\begin{array}{l}
r_{+}=r_1 \cos \vartheta+r_0, \\
\theta=r_2 \sin \vartheta+\frac{\pi}{2},
\end{array}\right.
\end{equation}
where $\vartheta\in (0,2\pi)$. In addition $r_{1}$ y $r_{2}$ re the parameters that determine the size of the contour to be drawn. Also,
$r_0$ is the point around which the contour is drawn. $r_1$, $r_2$ and $r_0$ are chosen in such a way that the contour $C$ encloses the defect or zero point of the vector field $n$. After that, the deflection of the vector field $n$ is found along the contour
$C$ as

\begin{equation}\label{defele}
\Omega(\vartheta)=\int_0^\vartheta \epsilon_{12} n^1 \partial_\vartheta n^2 d \vartheta,
\end{equation}
this is followed by the calculation of the winding number wi around the $i^{th}$ zero point of the vector field, as follows
\begin{equation}\label{win}
w=\frac{\Omega(2 \pi)}{2 \pi}.
\end{equation}
Finally, the topological charge $W$ can be determined by summing the winding numbers calculated along each
contour around the zero points, \i.e.
\begin{equation}
W=\sum_i w_i
\end{equation}
It is worth noting that when the parameter region does not have any zero points, the total topological charge is set
to zero.

\smallskip

\begin{figure*}[h!]
\centering
\includegraphics[width=0.45\textwidth]{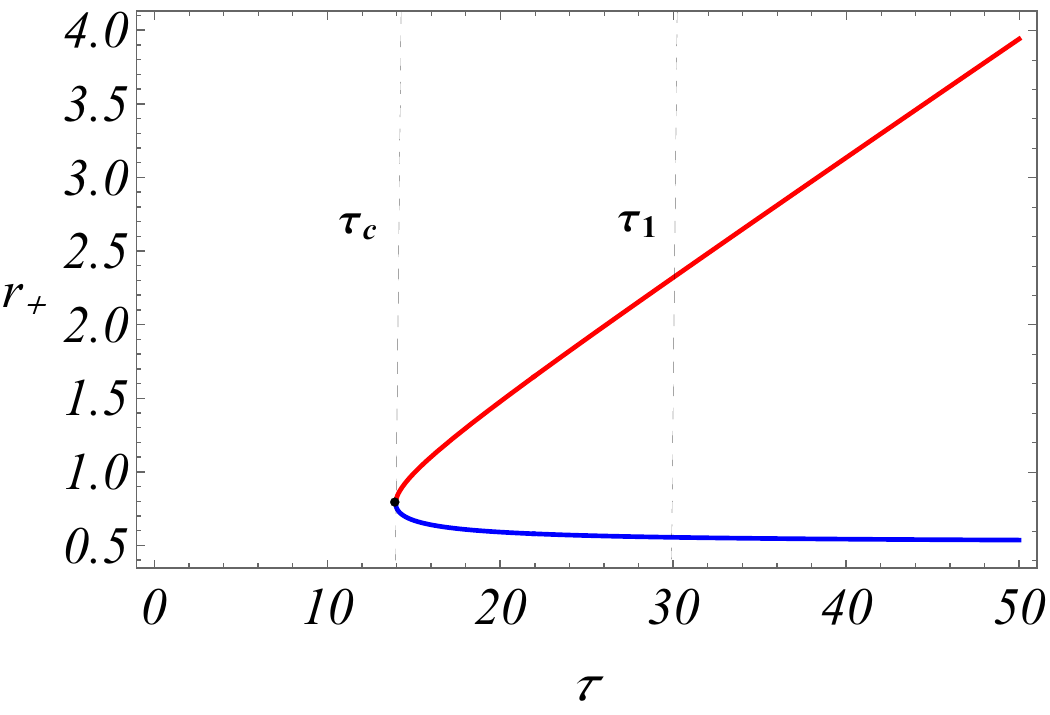}
\caption{Zero points of the vector $\phi$ shown in the $r_{+}-\tau$ plane.
To build this plot we have used $M$ 1[km] and $\alpha=0.02$ [$\text{km}^{2}$].
The black dot with $\tau$ given by (\ref{tau1}) denotes the generation point
for the black hole. At $\tau=\tau_{1}$, there are two  black holes.
\label{fig2}
}
\end{figure*}

\begin{figure*}[h!]
\centering

\includegraphics[width=0.32\textwidth]{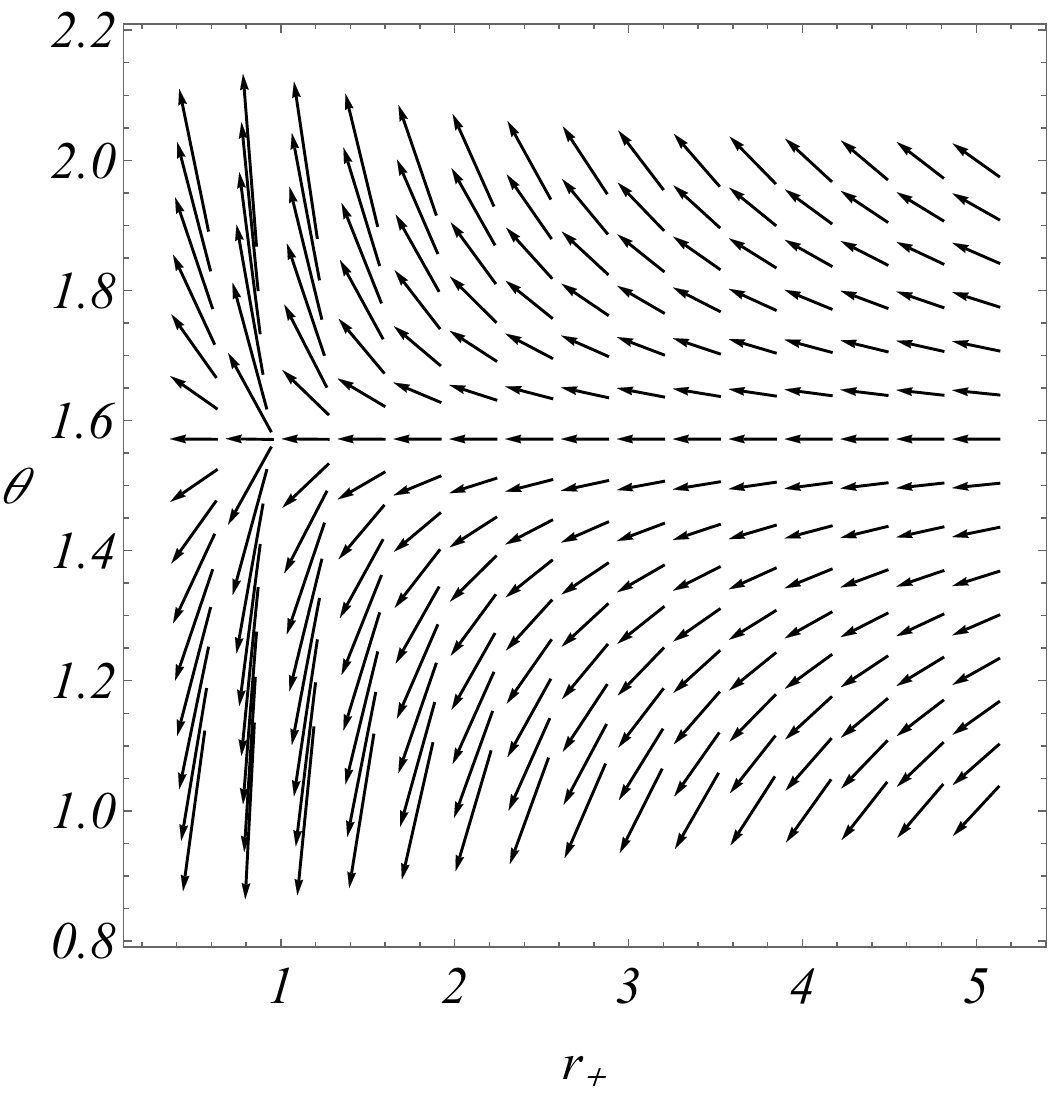}\ 
\includegraphics[width=0.32\textwidth]{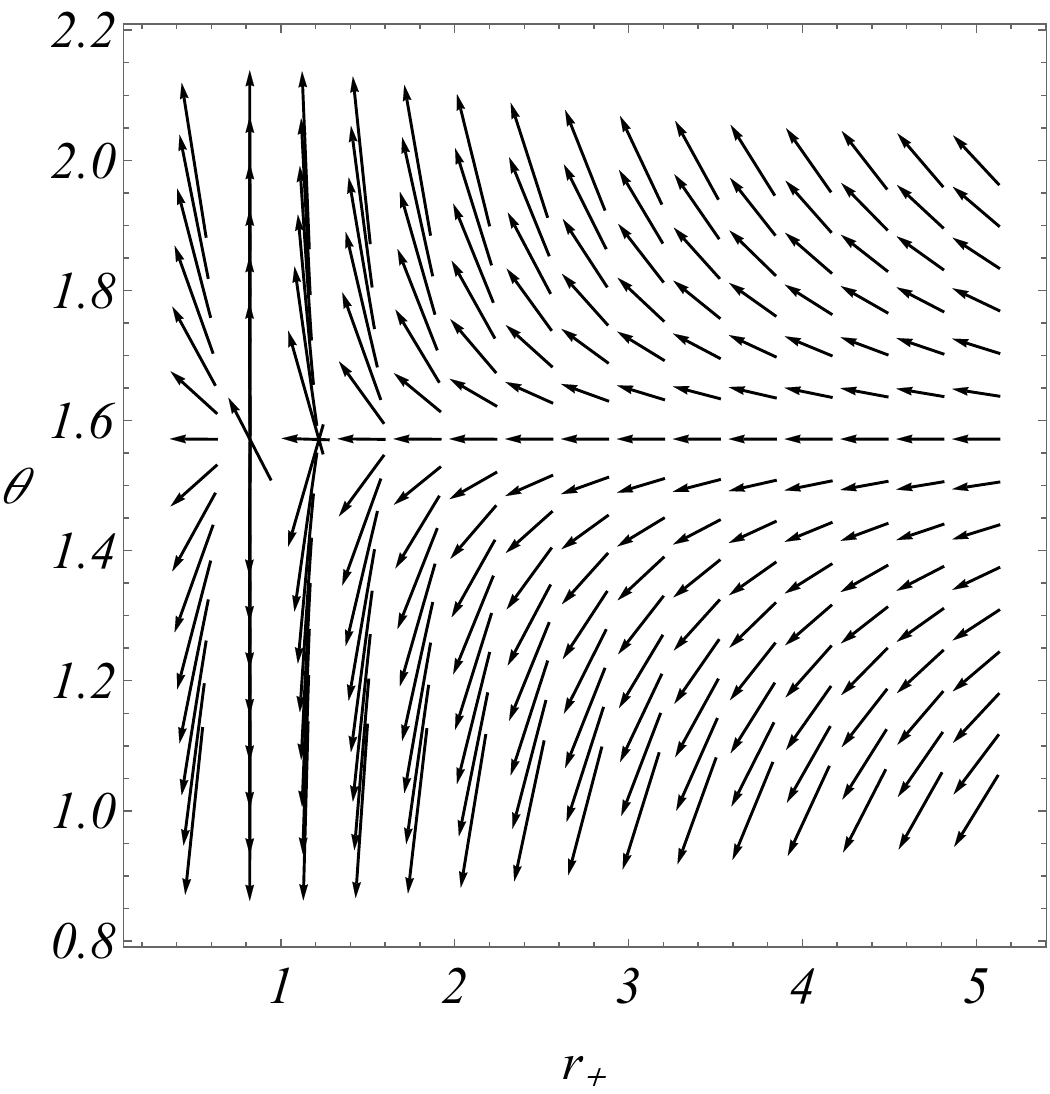}
\ 
\includegraphics[width=0.32\textwidth]{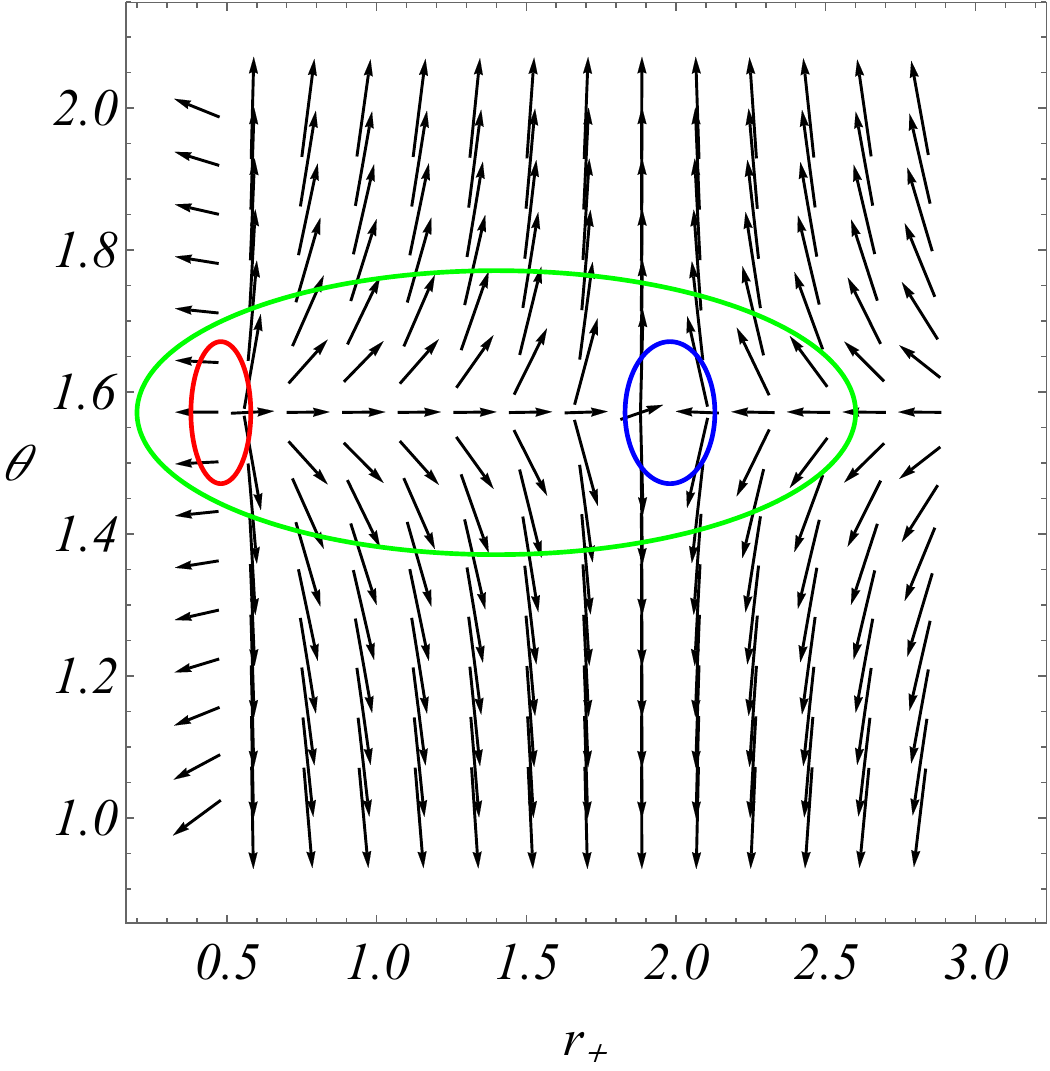}
\caption{The black arrows represent the unit vector field $n$ on a
portion of the $\theta-r_{+}$.
$(r_{+},\theta)=(0.7870,\pi/2)$ for the critical case (middle panel) and $(r_{+},\theta)=(0.5672,\pi/2)$ and $(r_{+},\theta)=(1.9049,\pi/2)$ (right panel) for $\tau>\tau_{c}$. The blue contours are
closed loops enclosing the zero points. For these plot we have chosen $M=1$, $\alpha=0.02$ and $\tau=60$.
\label{fig1}
}
\end{figure*}

\begin{figure}[h]
\centering
\includegraphics[width=0.4\textwidth]{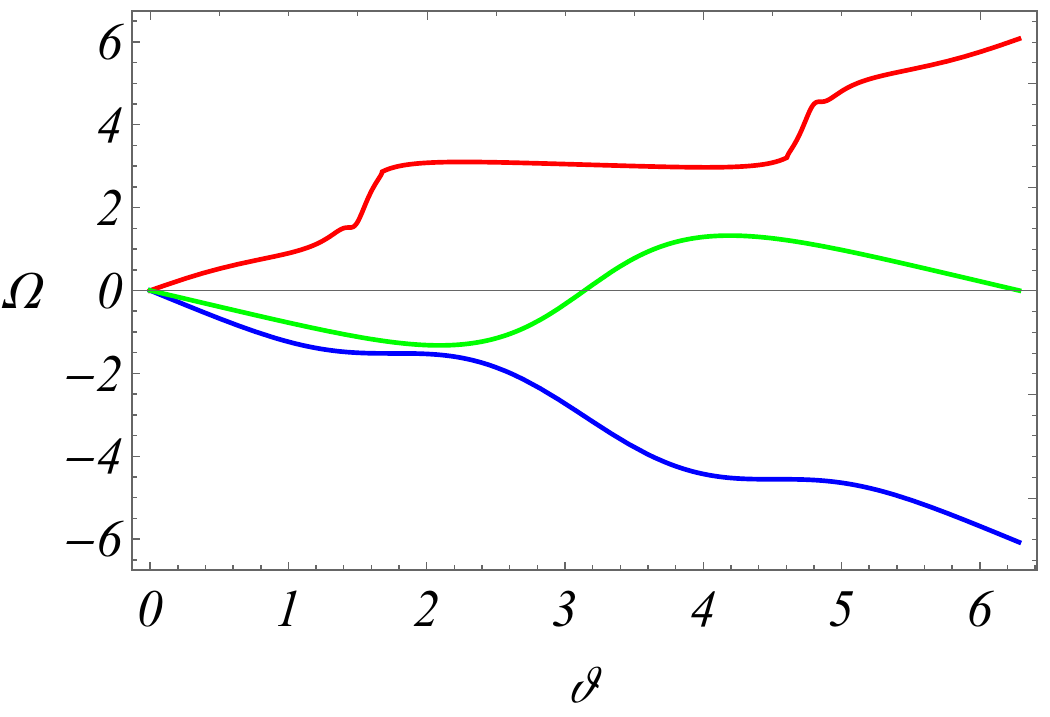}
\caption{The $\Omega$ vs $\vartheta$, for the contours exhibited in Fig. \ref{fig1} (right panel). For these plot we have chosen $\alpha=0.02$ and $\tau=25$.
\label{fig11}
}
\end{figure}

Now that the full picture is clear, we proceed in analyzing the topological thermodynamic behavior of the quantum BH model. However, to be consistent with the classical approach, we need to take into account that the horizon mass $M$ cannot replace the energy in the expression for the off-shell free energy $\mathcal{F}$. Instead be need to employ the expression (\ref{energyhorizon}), in this way the off-shell free energy for this model reads
\begin{equation}
    \mathcal{F}=E-\frac{S}{\tau}=\frac{1}{\alpha}\left[2\alpha r_{+} -r^{3}_{+}+(r^{2}_{+}-\alpha)^{3/2}\right]-\frac{\pi r^{2}_{+}}{\tau}. 
\end{equation}
So, using the definition (\ref{phi}), the components of the Duan's mapping are
\begin{equation}
\begin{aligned}
\phi^r & =2 + \frac{3 r (-r + \sqrt{r^2 - \alpha})}{\alpha} - \frac{2 \pi r}{\tau}, \\
\phi^{\theta} & =-\cot \theta \csc \theta .
\end{aligned}
\end{equation}
The unit vectors $(n^{1},n^{2})$
are computed using the equation (\ref{unit}). Now we calculate the zero points of the $\phi^{r}$
component by solving the following equation ($\phi^{r}=0$) and find an expression for $\tau$ in the following form
\begin{equation}\label{tau1}
  \tau= \frac{2 \pi r \alpha}{-3 r^2 + 3 r \sqrt{r^2 - \alpha} + 2 \alpha}.
\end{equation}
Next, we plot the horizon radius $r_+$ against $\tau$ in Fig. \ref{fig2}. We observe two branches of BHs. For large $\tau=\tau_{1}>\tau_{c}$ there are two intersection points. The intersection
points exactly satisfy the condition $\tau=T^{-1}$, and therefore denote
the on-shell BH solutions with temperature $\tau=T^{-1}$. For $\tau=\tau_{c}= 13.95$ (the critical value), the two intersection points for the BH
coincide and then disappear for $\tau<\tau_{c}$. Based on the local property
of a zero point, we have the topological number $W=1-1=0$. 
This fact is clear from the Fig. \ref{fig1}, where the vector plot is shown for the components of the unit vector $n^{a}$. The left panel shows the case below $t_{c}$, which does not exhibit any zero points of the vector $n$. The middle panel exhibits, the generating point located at $r_{+}=0.7870$ (the first zero point of the vector field). Finally, the right panel displays the zero points for $\tau>\tau_{c}$, of the vector field observed at $r_{+}= 0.5672$ and $r_{+}=1.9049$, respectively. The smallest one has a winding number $w=+1$ and the other other $w=-1$. Noting that each zero
point of the unit vector has a winding number 1 or -1,
from this, we conjecture that winding number is related
to local thermodynamic stability, with positive/negative
values corresponding to stable/unstable BH solutions. The red and blue contours are
closed loops enclosing the zero points (ZPs), while the green closed loop is enclosing both ZPs. The trend of the deflection of the vector field (\ref{win}), is displayed in Fig. \ref{fig11}. As can be observed, it gives $+1$ for the topological charge (\ref{win}) (red line corresponding to the red closed loop) and $-1$ (blue line, corresponding to the blue closed loop). These provides $W=0$ as shown by the green line (corresponding to the green closed loop). Thus this quantum BH system has the same
topology from the viewpoint of the thermodynamics as the Reissner-Nordstr\"om space-time.


\newpage

\section{Conclusions}\label{sec6}

In summary, in the present work we have studied a certain type of quantum BH that was recently proposed in the literature \cite{Lewandowski:2022zce}. The gravitational potential consists of the standard Newtonian one plus a correction that is proportional to a coupling $\alpha$ (see Eq. (\ref{newton})), the only free parameter. First, using observational data we put an upper bound on the coupling (see Fig. \ref{fig:1}). This analysis revealed that $\alpha$ can take greater values than those reported in \cite{Meissner:2004ju,Domagala:2004jt} and also, we found that this correction is not substantially affecting the Newtonian regime (see Eq. (\ref{correct})). However, we have chosen and small value than the reported in \cite{Meissner:2004ju,Domagala:2004jt} to see how this small correction is affecting the classical background.
After that, assuming a numerical value of $\alpha =0.02$ within its allowed range, we studied its QNMs for scalar and electromagnetic perturbations. Tables, \ref{table:First set}, \ref{table:Second set}, \ref{table:Third set} and \ref{table:Fourth set} show that this quantum BH is completely stable under scalar (massive and massless) and vector fields. Furthermore, we explore the stability from the thermodynamic point of view, using the classical approach \cite{Altamirano:2014tva} and the topological approach \cite{Wei:2021vdx,Wei:2022dzw}. These results show that after ``regularizing'' the thermodynamic first law, both approaches reveal that this quantum BH behaves as the Reissner-Nordstr\"om space-time. This means that this quantum BH has a stable and unstable branches, the small and large BH respectively (see right panel of Fig. \ref{fig}). In the topological approach, this fact is reflected on the existence of a generating point at $\tau_{c}$ (see Fig. \ref{fig2}). So, Duan's map displayed by Fig. \ref{fig1} (right panel), shows that the unstable branch is represented by the blue closed loop with topological charge equal to $-1$, and the stable branch corresponds to the one enclosed by the red loop, providing a topological charge $+1$. Therefore, the global charge for this model is $W=0$. Finally, we can say that the quantum BH model is stable under mechanical perturbations, but its thermodynamical stability depends on the values of the parameter space $\{M;r_{+};\alpha\}$, which determines the stable branch.

\bibliography{biblio.bib}
\bibliographystyle{elsarticle-num}

\end{document}